\documentclass[10pt,letterpaper,english,superscriptaddress,nofootinbib,pra,twocolumn,notitlepage]{revtex4-1}

\usepackage{amsmath}
\usepackage[matrix,frame,arrow]{xypic}
\usepackage{amscd}
\usepackage{amssymb}
\usepackage{graphicx}

\usepackage[colorlinks = true, linkcolor=ddgreen, urlcolor = ddgreen,citecolor=ddgreen,pdfpagemode = None]{hyperref}

\newcommand{\figref}[1]{Fig.\;\ref{#1}}
\newcommand{\secref}[1]{Sec.~\ref{#1}}
\newcommand{\appref}[1]{App.~\ref{#1}}
\newcommand{\Ssref}[1]{\S\ref{#1}}
\newcommand{\Eeqref}[1]{Eq.~\eqref{#1}}
\newcommand{\refcite}[1]{Ref.~\cite{#1}}
\newcommand{\abs}[1]{\left\vert#1\right\vert}
\newcommand{\set}[1]{\left\{#1\right\}}
\newcommand{\ket}[1]{\left\vert#1\right\rangle}
\newcommand{\nket}[1]{\vert#1\rangle}
\newcommand{\bra}[1]{\left\langle#1\right\vert}
\newcommand{\kb}[2]{\vert#1\rangle\!\langle#2\vert}
\newcommand{\bk}[2]{\langle #1\vert#2\rangle}
\newcommand{\Bigbk}[2]{\Big\langle #1\Big\vert#2\Big\rangle}
\newcommand{\ex}[1]{\langle#1\rangle}
\newcommand{\proj}[1]{\kb{#1}{#1}}

\newcommand{\half}{\frac12}
\newcommand{\ot}{\!\otimes\!}
\newcommand{\di}{\mathrm{d}}
\newcommand{\ii}{\mathrm{i}}
\newcommand{\Ee}{\mathrm{e}}

\newcommand{\m}[1]{\mathcal{#1}}

\newcommand{\Hi}{\mathcal{H}}

\newcommand{\M}{\mathcal{M}}

\DeclareMathOperator{\Tr}{Tr}

\usepackage{color}
\definecolor{ddgreen}{rgb}{0,0.4,0}

\begin{document}

\title{Changing quantum reference frames}
\author{Matthew C. Palmer}
\affiliation{Centre for Engineered Quantum Systems, School of Physics, The University of Sydney, Sydney, NSW 2006, Australia}
\author{Florian Girelli}
\affiliation{Department of Applied Mathematics, University of Waterloo, Waterloo, Ontario, Canada}
\author{Stephen D. Bartlett}
\affiliation{Centre for Engineered Quantum Systems, School of Physics, The University of Sydney, Sydney, NSW 2006, Australia}
\date{2 May 2014}

\begin{abstract}
We consider the process of changing reference frames in the case where the reference frames are quantum systems.  We find that, as part of this process, decoherence is necessarily induced on any quantum system described relative to these frames.  We explore this process with examples involving reference frames for phase and orientation.  Quantifying the effect of changing quantum reference frames provides a theoretical description for this process in quantum experiments, and serves as a first step in developing a relativity principle for theories in which all objects including reference frames are necessarily quantum.
\end{abstract}

\maketitle

\section{Introduction\label{sec-intro}}

Quantum states and dynamics are commonly described with respect to a classical background reference frame. Even defining a basis for the Hilbert space of a quantum systems will in general make reference to a background frame. For example, the state $\ket 0$ for a spin--$\half$ particle may be defined as the spin parallel to the $z$-axis of a laboratory reference frame, and the $(\ket0+\ket1)/\sqrt2$ state as spin parallel to the $x$-axis. In place of a classical background frame, one could use a second quantum system prepared in a state that indicates an orientation or alignment of a frame. The basis can then be defined with respect to this `quantum reference frame.' In this case, quantum information is encoded in degrees of freedom that are independent of the orientation of the laboratory reference frame.

In this paper we consider the process of changing quantum reference frames.  That is, we begin with a situation wherein the quantum state of a system $S$ is defined with respect to a quantum reference frame $A$, and we seek a procedure such that in the end we are describing the same quantum system with respect to a new quantum reference frame $B$.  We will focus on the scenario in which we have no prior knowledge of the orientation of the new quantum reference frame $B$ with respect to the original frame $A$. As a concrete example, consider a quantum optics experiment where the quantum state of an atom is defined with respect to a phase reference in the form of a laser, denoted $A$. We want to switch to a new phase reference $B$, i.e., a different laser, for which the phase relationship between $A$ and $B$ is initially unknown.  In order to describe our atom with respect to laser $B$, some procedure must be performed that correlates the two lasers, ideally in a way that minimally affects the quantum state of the atom; practically, this may involve phase locking the lasers \cite{BRSDialogue}.  (Note that there are other possible ways of defining a `change of quantum reference frame,' for example, when there are two reference frames for which the observer knows the orientation, changing reference frames is simply a matter of discarding the undesired reference frame.  Another interpretation of a `change' is a single reference frame transformed under a symmetry action.)  In this paper, we will investigate and quantify how the description of the quantum system $S$ changes as a result of this change of quantum reference frame process. This analysis provides the theoretical description of a process that occurs in numerous experimental guises \cite{BRSDialogue,PfleegotMandel67PR,PfleegorMandel67PLA,Molmer97PRA,CKR05,RudolphSanders01,Fujii03, JavanainenYoo96,HostonYou96,WCW96,ATMDKK97,CastinDalibard97,CKR05,BRSDialogue,DB00,Denschlag00},  and may also form the first element of a relativity principle for quantum reference frames, which dictates how the description and dynamics of a physical system changes under a change of reference frame.

Quantum reference frames in general will use finite resources, quantified by some parameter such as the Hilbert space dimension of the frame.  If our reference frames describe a continuum of orientations and we restrict the size (e.g.\;Hilbert space dimension) of the frames, then reference frame states corresponding to different orientations will not be perfectly distinguishable.  This uncertainty of the frame results in decoherence in information encoded using the reference frame~\cite{BRST}. In particular, as we will show, decoherence can result from a change of quantum reference frame. This decoherence would be important to limit in quantum experiments, and would also be a novel effect for a relativity principle. With a construction and characterisation of a `change of quantum reference frame' procedure, we quantitatively investigate the decoherence resulting from changing physical quantum reference frames.  We interpret this appearance of decoherence in terms of a type of intrinsic decoherence, which is a proposed semiclassical phenomenon of quantum gravity arising from fundamental uncertainties in the background space  \cite{Power00,KokYurtsever,Gambini04,Milburn03,GirelliPoulin07,GirelliPoulin08}.

The structure of the paper is as follows. \secref{sec:Prelim} reviews the concepts and mathematical formalism of quantum reference frames. \secref{sec-introCoRF} presents the definition of the quantum operation describing a change of quantum reference frame, and an analysis of the properties of this quantum operation for some special cases of quantum reference frame.  We also discuss the significance of the decoherence induced, and what consequences the procedure has. In Secs.~\ref{sec-phaseex} and \ref{sec-exampleSU2} we provide examples of the procedure for phase references (characterised by the group $U(1)$), and a Cartesian frame and direction indicator (characterised by $SU(2)$). In \secref{sec-conclusion} we present some concluding remarks.

\section{Preliminaries:  Classical and quantum reference frames\label{sec:Prelim}}
In this section, we review the conceptual and mathematical tools used in the description of a quantum state relative to a classical or quantum reference frame.  We follow the notation of the review article~\cite{BRS}.

\subsection{Types of reference frames}
Physical quantities are often defined (perhaps implicitly) with respect to a non-dynamical background  reference frame, e.g.\;a coordinate system. We want to use physical objects as reference frames and define quantities with respect to these. In this paper we consider the quantum case in which the physical objects are quantum systems which will be used as reference frames for other quantum systems. These `quantum reference frames' are therefore physical systems with their own dynamics, since they are included in the description just as any other quantum system. When using a physical system as a reference frame, well-defined relationships between the frame system and other systems can be meaningful even without a background reference frame.  Such quantities are called \emph{relational degrees of freedom}, which we define in detail below.

We also distinguish between a reference frame that has some correlation with the system of interest, which we call an \emph{implicated} reference frame, and one that is completely uncorrelated with the system, which we call a \emph{nonimplicated} reference frame.  Consider the example of the quantum optics experiment described in the introduction. An experiment is performed involving the state of an atom defined relative to a phase reference laser. This phase reference is \emph{implicated} with the system of atoms. Another independent laser, not phase locked with the first, is a nonimplicated frame for the system.

\subsection{Quantum reference frames}

A quantum reference frame is a quantum system and is therefore described by a state in an associated Hilbert space. Different quantum states in this Hilbert space can describe different `orientations' of the quantum reference frame with respect to a hypothetical background frame. To formalise these notions, we will look at how to mathematically describe the manipulation of quantum reference frames and relational quantum degrees of freedom using techniques from the theory of group representations.

To set up the use of a quantum reference frame for encoding information in relational degrees of freedom, we begin with a background reference frame and a quantum system $S$ in the state $\rho_S$ with respect to this background frame.  Changes of orientation of this system relative to the background frame are described by a unitary representation $U_S(g)$ of an element $g$ from a group $G$ which describes all possible changes of orientation.

Next, consider an additional system $R$ prepared in a quantum reference frame state $\rho_{R}$, also defined with respect to the background reference frame. The reference frame state breaks a symmetry associated with $G$, which has a representation as a unitary $U_R$ acting on the Hilbert space of the quantum reference frame system $\Hi_R$. We can now consider using $R$ as a reference frame for $S$.

To ensure that we are not still making accidental use of the background frame, we de-implicate it. This de-implication involves decorrelating the compound quantum system $SR$ from the background frame. For a general state $\rho$, this is done by averaging the state over all rotations $g\in G$ using $U_{SR}(g) = U_S(g) \otimes U_R(g)$, the unitary representation of $G$ on the combined (tensor product) Hilbert space of $S$ and $R$. The resulting map $\m G$ is called the \emph{G-twirl} of the state, given by
\begin{equation}
\m G\left(\m \rho\right)=\int\di\mu(g)\m U_{SR}(g)[\rho] = \int\di\mu(g) U_{SR}(g) \rho U_{SR}(g)^\dag\label{eq-Gtwirl}
\end{equation}
where $\m U_{SR}(g)[\rho]:=U_{SR}(g)\rho U_{SR}(g)^\dag$ is the unitary map of the left action of the group and $\di\mu(g)$ is the group-invariant Haar measure of the group (for example, the $U(1)$ integration measure is $\di\theta/2\pi$). (Note that although here we are considering $U_{SR}(g)$, this G-twirl map can be generalised to any unitary representation.) We restrict our attention to compact Lie groups, where the average is well-defined and bounded, although we note there exist similar methods for more general cases.  We will call the states that are invariant under G-twirling $\sigma=\m G(\sigma)$ (including G-twirled states $\sigma=\m G(\rho)$) `group-invariant' or `G-invariant'. These states are well-defined independent of a background reference frame. Note that a $G$-twirled state may be mixed even if the original state $\rho$ was pure.

\subsection{Relational degrees of freedom}

Relational degrees of freedom are those which are independent of any background frame. Given a system state $\rho_S$ and a quantum reference frame $\rho_R$, it is not immediately obvious what are the relational degrees of freedom in the $G$-twirled joint state $\m G(\rho_S \otimes \rho_R)$. In the following, we will define the Hilbert space subsystems associated with these relational degrees of freedom, following~\cite{BRST,BRS}.  Again, for simplicity of the mathematics, we will consider symmetries corresponding to compact Lie groups such as $U(1)$ and $SU(2)$. However, many of the concepts developed can be directly transferred to general groups and reference frames.

The unitary representation of a compact Lie group on a Hilbert space $\Hi$ consists of a number of inequivalent representations called `charge sectors'. The Hilbert space can be decomposed into a tensor sum of these charge sectors, each labelled by $q$ (for example, $q$ may be total spin in a representation of $SU(2)$ on a collection of spins). Each of these charge sectors may be a \emph{reducible} representation, which can be further decomposed into a Hilbert subsystem $\m M^{(q)}$ carrying an \emph{irreducible} representation (`irrep'), and a `multiplicity subsystem'  $\m N^{(q)}$ which carries the trivial representation and whose dimension indicates how many copies of the irreducible representation exists in the charge sector $q$. The representation on the full Hilbert space then has the structure
\begin{equation}
\Hi=\bigoplus_{q} \m M^{(q)}\otimes\m N^{(q)},\label{eq-repdecomp}
\end{equation}
where $q$ ranges over all the irreps (charge sectors) of $G$ that are supported on $\Hi$.

The $G$-twirl map \eqref{eq-Gtwirl} is closely related to the representations of the group, in that it averages an input state $\rho$ over the unitary action of every element in the symmetry group. Decomposing this map following \eqref{eq-repdecomp}, we have
\begin{equation}
\m G(\rho)=\sum_{q}(\m D_{\m M^{(q)}}\otimes\m I_{\m N^{(q)}})[\Pi^{(q)}\rho{\Pi^{(q)}}^\dag].\label{eq-Gtwirldecomp}
\end{equation}
The terms in this operation are defined as follows.  First, $\Pi^{(q)}$ is the projector onto the subspace $\m M^{(q)}\otimes\m N^{(q)}$, the charge sector $q$. This removes all coherences between the charge sectors. Next, $\m D$ is the completely depolarising channel, which is a trace preserving map that takes every density operator to a scalar multiple of the identity operator on the space $\m M^{(q)}$. This is the effect of an average of the action of a unitary group on an irrep. Finally $\m I^{(q)}$ is the identity map on the multiplicity subsystem $\m N^{(q)}$.

We can now identify the relational degrees of freedom, unaffected by $G$-twirl, as the multiplicity subsystem $\m N^{(q)}$.  The degrees of freedom in the subsystems $\m M^{(q)}$ are defined only with respect to a background frame, and are completely decohered by the $G$-twirl.

\subsection{Quantum reference frame states}\label{sec-groupestates}\label{sec-RFquant}

In this section we describe how to define useful quantum reference frame states. A reference frame breaks a symmetry by indicating an orientation.  The set of possible orientations is associated with a symmetry group $G$. To construct a set of reference states, we begin with a \emph{fiducial state} $\ket{\psi(e)}$, which serves as a quantum reference frame oriented with respect to a background frame and which we choose to associate with the identity $e\in G$. Given this fiducial state we can construct states corresponding to other orientations $g\in G$ by generating the states in the orbit of $\ket{\psi(e)}$ under the group action $U(g)$, yielding $\ket{\psi(g)}:=U(g)\ket{\psi(e)}$ for all $g\in G$.  Such states obey the relation $U(h)\ket{\psi(g)} = \ket{\psi(hg)}$, and we say that they transform \emph{covariantly} under the action of the symmetry group.

Quantum reference frames generally use limited finite resources quantified by some parameter $s_{R}$. A fundamental example of a size parameter is the dimensionality of the Hilbert space $\Hi_{R}$, constraining the number of charge sectors $q_{R}$ under the representation of the group (however, this is not the only choice of size parameter). We define the notation $\ket{s_R;\psi(g)}$ to denote a $G$-covariant state $\psi(g)$ with size parameter $s_R$.  Where it is unnecessary to indicate size, we may suppress the size parameter. The groups considered in the theory of this paper are compact Lie groups, meaning the reference frames can take one of a continuum of orientations in a closed manifold. With only finite-dimensional representations of such groups, reference frame states for different orientations in a Lie group cannot all be perfectly distinguishable. Consequently, a state will have a mean orientation $g$, but also possess an uncertainty in orientation.

We would like reference frame states to have a well-defined classical limit in which the overlap of states with different orientations becomes zero as the size parameter $s_{R}$ increases to infinity, i.e.,
\begin{equation}
  \lim_{s_R \to \infty} D_{s_R}\abs{\bk{s_R;\psi(g)}{s_R;\psi(h)}}^2 = \delta(gh^{-1}),
\end{equation}
where $\delta(g)$ is the delta function on $G$ defined by $\int\di\mu(g)\delta(g)f(g)=f(e)$ for any continuous function $f$ of $G$ \cite{BRST}, and $D_{s_R}$ is the dimension of the Hilbert space spanned by $\{\ket{s_R;\psi(h)} ; h\in G\}$.

In the finite size case, one may wish to maximise the distinguishability of the quantum reference frame used for a given size constraint $s_R$. Distinguishability can be quantified using maximum likelihood or fidelity measures \cite{Chiribella06,BRS,Bisio10,MarvianSpekkens13,GMS09}. Because we also want the reference frame states to become ideal (perfectly distinguishable) in the classical limit, we want this distinguishability to scale with $D_{s_R}$ (see \cite{Chiribella06,Vaccaro08,MarvianSpekkens13,GMS09} regarding asymptotic measures). A useful choice of reference frame states for a group $G$ on $D_{s_R}$ dimensions are the \emph{maximum likelihood states} \cite{Chiribella06}, denoted $\ket g$ or $\ket{s_R;g}$ (the latter following the notation $\ket{s_R;\psi(g)}$), as these states are optimal for a range of operational tasks involving reference frames. These pure states transform covariantly and have the property that
\begin{equation}
  \m G(|g\rangle\langle g|) = D_{s_R}^{-1}I\,,\label{eq:uniformGtwirl}
\end{equation}
i.e.\;these have uniform support over their Hilbert space, which will make these states useful in the construction of measurements (POVMs). The form of a maximum likelihood state is specific to the group $G$ and Hilbert space $\Hi_R$. We will consider quantum reference frame Hilbert spaces $\Hi_R$ for which the decomposition \eqref{eq-repdecomp} of just $\Hi_R$ is such that $d_q:=\dim\m N^{(q)}=\dim\m M^{(q)}$ for every charge sector $q$ \cite{BRST}. Then the maximum likelihood states take the simple form \cite{Chiribella04}
\begin{equation}
\ket e=D_{s_R}^{-\half}\sum_{q\in Q_R}d_q\ket{\Psi^{(q)}}
 \end{equation}
where in this case $D_{s_R}=\sum_{q}d_q^2$ and $\ket{\Psi^{(q)}}=d_q^{-\half}\sum_{m=1}^{d_q}\nket{\phi_m^{(q)}}\otimes\nket{r_m^{(q)}}$ is a normalised maximally entangled state on $\m M^{(q)}\otimes\m N^{(q)}$ in some pair of bases $\nket{\phi_m^{q}}\in\m M^{(q)}$ and $\nket{r_m^{(q)}}\in\m N^{(q)}$. The set $Q_R$ includes the charge sectors on which this state has support and is determined by the size parameter $s_R$. It is straightforward to generalise the machinery to cases where $\dim\m M^{(q)}\neq\dim\m N^{(q)}$; see Refs.~\cite{BRST,BRS,Chiribella04,Chiribella06,CDMPS04,KMP04}. For more details regarding properties of these states see \refcite{BRST,Bagan01,PeresScudo01,Bagan04,Chiribella05,Lindner06,Bagan06,BRS}. We will be using examples of maximum likelihood states as quantum reference frames in Secs.~\ref{sec-phaseex} and \ref{sec-exampleSU2}.

\subsection{Encoding and recovering relational states\label{sec-quantRF}\label{sec-introdequant}\label{sec-REOptimal}}

One essential task when using quantum reference frames is to implicate a quantum reference frame for a system state $\rho_S$ while de-implicating the background frame, called \emph{relationally encoding} $\rho_S$. A second essential task is to do the reverse: extract the information from this encoding by removing the quantum reference frame and \emph{recovering} a $\rho_S'$ defined relative to a background frame. In this section we define the operations that do these tasks.

Given a quantum system $S$ defined relative to a background frame, we want to introduce a quantum reference frame $R$ and de-implicate the background frame. This is achieved by the \emph{encoding map}
\begin{equation}
\m E_{\rho_R}(\rho_S):=\m G_{S R}\left(\m \rho_S\otimes \rho_{R}\right).\label{eq-encGC}
 \end{equation}
\Eeqref{eq-encGC} results in a G-invariant state $\sigma_{SR}=\m E_{\rho_R}(\rho_S)$ which is called the relational encoding of $\rho_S$ using $\rho_R$. The map can be implemented by applying \eqref{eq-Gtwirl} to $\rho_S\otimes\rho_{R}$, where the unitary representation on the compound Hilbert space $\Hi_S\otimes\Hi_R$ is given by $U_S(g)\otimes U_{R}(g)$. It can also be implemented by the projection to charge sectors and depolarisation of irreps given by \eqref{eq-Gtwirldecomp}.

Now, if we are given an encoded state but wish to recover the state $\rho_S$ in terms of a background frame, this usually cannot be done perfectly \cite{BRST}. The procedure we use is the \emph{recovery map}~\cite{BRST}
\begin{multline}
\m R(\sigma_{S R})=\\D_{s_R}\int\di\mu(g)\bigl[U_S(g^{-1})\otimes\bra{g}_{R}\bigr]\sigma_{S R}\bigl[U_S(g^{-1})^\dag\otimes\ket{g}_{R}\bigr],\label{eq-rec}
\end{multline}
which results in a state $\rho_S'$ on $\Hi_S$. This map describes the measurement of the quantum reference frame on system $R$ against a background reference frame, described by a covariant POVM formed with elements proportional to projectors onto the states $\ket g_R$ on the reference frame. If the reference frame is measured to have orientation `$g$' relative to the background frame, then the orientation of the state is corrected by a transformation by $g^{-1}$.

The finite size of a quantum reference token means that for symmetries described by compact Lie groups, the token is an imperfect reference frame. Consequently, the use of a quantum reference frame causes an effective decoherence to the information in $\rho_S$. We can describe this decoherence by composing \eqref{eq-encGC} and \eqref{eq-rec} to produce
\begin{equation}
\m R\circ\m E_{\rho_R}(\rho_S)=D_{s_R}\int\di\mu(g)\bra g\!\rho_R\!\ket g\m U_S(g^{-1})[\rho_S].\label{eq-REmap}
\end{equation}
This map takes the form of a noise map on $\rho_S$, describing a mixing of this state over a distribution of unitaries determined by the distribution $\bra g\rho_R\ket g$. We want to minimise this decoherence by optimising the recovery operation to produce the state closest to $\rho_S$ possible from an encoding $\m E_{\rho_R}(\rho_S)$. The figure of merit used to quantify this optimisation is the average entanglement fidelity of an arbitrary input ensemble of states $\sigma_{SR}$ into $\m R$ \cite{BRST}. The recovery map is generically near-optimal in the sense that if the average fidelity of an optimal recovery map $\m R_\text{opt.}$ is $\bar F_e=1-\eta$ then the recovery map $\m R$ has average fidelity $\bar F_e\ge(1-\eta)^2\ge1-2\eta$, i.e.\;the error $\eta$ is never greater than twice that of the best recovery operation \cite{BarnumKrill02}.

\section{Change of a quantum reference frame}
\label{sec-introCoRF}

We now consider the central problem of changing quantum reference frames. We begin this section with a qualitative discussion of the issues regarding measurement when changing reference frames, including an example to illustrate the central ideas. If the reader prefers, this subsection \Ssref{sec-introchangingRFs} can be skipped in favour of the mathematical formulation in \Ssref{sec-genresults}.

\subsection{Changing quantum reference frames:  a qualitative discussion\label{sec-introchangingRFs}}

As an example, also investigated in \cite{AharonovKaufherr,ABPSS}, consider a particle $S$ in one dimension, with position defined relative to a reference frame consisting of another particle $A$ which provides an origin.  Introduce a second particle, $B$, which we would like to use as a new reference frame for $S$.  Classically, this seems straightforward:  the position of $S$ described in terms of $B$ will differ by the relative position of the two reference frames, $x_B-x_A$.  (Note that this relational quantity is independent of any choice of origin.) After adjusting our description of the position of $S$ by this difference, particle $A$ can be subsequently discarded.

In this classical scenario, we can implicitly assume that the relative position of the two frames, $x_B-x_A$, is known a priori.  In the quantum scenario we consider, the reference frame $B$ is initially deimplicated, meaning that it is uncorrelated with either $A$ or $S$; in general we would require a measurement to determine such relationships.  There are two natural options for doing this. The relationship between $S$ and $B$ can be directly measured, or the relationship between $A$ and $B$ can be measured (giving us the relational quantity $x_B-x_A$ for adjusting the description of $S$). Let us concentrate on the quantum mechanical case now, and first consider a semiclassical configuration in which the $A$ and $B$ reference frames are in position eigenstates and the measurements are ideal projective measurements of relative position. The $S$ state is arbitrary. For the first measurement option, the relative position of $S$ and $B$ is measured. The wavefunction of $S$ will in general not be a position eigenstate, and therefore will be disturbed by a measurement of relative position; specifically, a projective measurement of $x_B-x_S$ in the situation where $B$ is in a position eigenstate will collapse the wavefunction of $S$ to a position eigenstate as well.  This complete disturbance is not consistent with what we expect of a change of reference frame. Instead, consider an alternative, where the relative position $x_B-x_A$ of the two reference frames $A$ and $B$ is measured, and the system $S$ is not involved. After obtaining a well-defined value of $x_B-x_A$, we can combine this with preexisting correlation between $S$ and $A$ to obtain a correlation between $S$ and $B$, since the associated operators for $x_S-x_A$ and $x_S-x_B$ commute.  This act of measurement has implicated the reference frame $B$, and we can now discard $A$. The new description of the state $S$ will have changed by $x_B-x_A$ due to the difference in position of reference frame $B$ versus $A$, thus accomplishing a change of quantum reference frame.

There are however some subtleties in this procedure. In the above example $A$ and $B$ were position eigenstates and the measurements were projective to these position eigenstates, allowing for arbitrarily good precision in the relational variables.  In general, quantum reference frames for generic degrees of freedom will not possess such ideal, perfectly-distinguishable configurations due to their bounded size~\cite{BRST}.  We will see that the bounded nature of the $A$ reference frame results in decoherence to the quantum system $\rho_S$ after $A$ is discarded. If the measurement is also only capable of projecting $x_B-x_A$ to a state with finite variance in position, then discarding the frame $A$ yields a system wavefunction correlated with an imperfect reference frame $B$. Recovering $\rho_S$ from quantum frame $B$ will then also cause decoherence to the system.

In the next section, we will formalise these concepts and problems, and construct a general framework for describing a change of quantum reference frame.  In particular, because we use quantum states to indicate orientations in a continuous group, in many cases we cannot perfectly distinguish nonorthogonal states for different reference frame orientations. One of the main limiting factors for distinguishability is the dimension of the Hilbert space used for the reference frame. The imperfect distinguishability results in an uncertainty in the orientation given by a quantum reference frame, leading to decoherence when we change the quantum frame used for encoding a quantum system.

\subsection{General results of change of quantum reference frame procedure\label{sec-genresults}}

In this section, we formally develop the change of quantum reference frame procedure and then calculate the final states for a physically relevant class of initial states.

We can formulate the notion of changing quantum reference frames as an operational task.  An observer possesses a quantum system $S$ and implicated quantum reference frame $A$; the initial state of this combined system is given by the encoding $\m E_{\rho_A}(\rho_S)$ given in \Eeqref{eq-encGC}. This observer wishes to use a second, initially non-implicated, quantum reference frame $B$. The task of the observer is to use the $B$ quantum system as a quantum reference frame for the system $S$, and to discard the initial reference frame $A$.  That is, the observer seeks to end up with a new encoding $\m E_{\rho_B}(\rho_S')$ of the system $S$ with respect to $B$, where we note that the state of the system $\rho_S'$ may have changed as a result of this procedure.  We measure success at this operational task by determining how close (relative to some natural figure of merit, such as fidelity or trace distance) is the final encoded state $\rho_S'$ compared with the original $\rho_S$.

As well as being of theoretical interest, this occurs in several experimental guises \cite{BRSDialogue}. For example, in switching phase or clock reference lasers from a locked laser $A$ to an uncorrelated laser $B$, one first needs to phase lock the two lasers \cite{PfleegotMandel67PR,PfleegorMandel67PLA,Molmer97PRA,CKR05} (this has been extended to issues in optical teleportation \cite{RudolphSanders01,Fujii03}). In another example, determining the relative phase of two Bose--Einstein condensates \cite{JavanainenYoo96,HostonYou96,WCW96,ATMDKK97,CastinDalibard97,CKR05} can be interpreted as correlating quantum reference frames \cite{BRSDialogue,DB00,Denschlag00}.

In the following sections, we develop such a procedure based on a relational measurement between the old and new quantum reference frames, $A$ and $B$.  We then consider how the encoded state on $S$ is affected by this procedure, and quantify how well this procedure performs at the above operational task.

\subsubsection{A measurement to determine the relationship between frames\label{sec-demoprop}}

The core element of the procedure to change quantum reference frames is a relational measurement of the two reference frames $A$ and $B$ that determines a relative orientation $h \in G$ between these two frames~\cite{BRSrelational}.  Performing this measurement leads to a correlation in orientation of the two reference frames. Because there was initially correlation between frame $S$ and the system $A$, we obtain correlation between $S$ and $B$. Now we can discard the $A$ reference frame by tracing and use $B$ as the new quantum reference frame. If the reference frames use finite resources such as finite Hilbert space dimension to indicate orientations, we expect decoherence in the post-measurement state.  The fact that it is a relational measurement means that it can be made independent of any background reference frame. In the following, we construct the relational POVM and update map for this measurement, and prove key properties of the construction.

The quantum statistics of a relational measurement of the two reference frames $A$ and $B$ are given by a relational POVM $\{E_h|h\in G\}$.  A POVM allows us to calculate the probabilities of the $h$ outcomes for an input state, but here we are equally interested in the post-measurement state.  We therefore construct a family of trace-decreasing completely positive (CP) maps $\m M^h_{AB}$ associated with the POVM elements to determine the post measurement state for a given outcome $h$.  (Such maps, which describe the POVM and also the post-measurement update rule, are sometimes called \emph{instruments}.)  We require these operations to be implementable without the use of a background reference frame.

We now define a measurement, as a POVM, satisfying the above conditions. The POVM is designed to determine orientation within the symmetry group, so will be formed from the maximum likelihood states $\{ \ket g; g\in G\}$ for the particular symmetry group $G$ of the scenario, using the techniques of \secref{sec-groupestates}.\footnote{It was shown in \cite{CDMPS04} that this is the POVM for measuring the orientation of a quantum reference frame which maximises the likelihood. Note that the recovery map \eqref{eq-rec} also uses a POVM of this form.} The maximum likelihood states $\ket g_A$ and $\ket g_B$ for each reference frame system $A$ and $B$ satisfy the conditions $\m G(|g\rangle_A \langle g|) = D_{s_A}^{-1} I_A$ and $\m G(|g\rangle_B \langle g|) = D_{s_B}^{-1} I_B$, where $D_{s_*}$ are normalisation factors given by the dimensions of the Hilbert space spanned by each projector on $A$ and $B$. We construct a family of projectors $\Pi_{AB}^{g,h}$ on the two reference frame systems $AB$ given by
\begin{equation}
\Pi_{AB}^{g,h}= \proj g_A\ot \proj{gh}_B=\m U_{AB}(g)[\proj e_A\ot\proj h_B],\label{eq-proj}
\end{equation}
with $\ket g=U(g)\ket e$. The projector $\Pi^{g,h}_{AB}$ projects onto the state describing an orientation $g\in G$ of the state on $A$ and an orientation $gh\in G$ of the state on $B$.

The projectors are defined with respect to a background frame.  By using a $G$-twirl, we can define relational POVM effects $\{E_h\}$ as
\begin{equation}
E_h=D_{s_A}D_{s_B}\int \di\mu(g)\Pi_{AB}^{g,h}.\label{eq-MPOVME}
\end{equation}
This measurement satisfies POVM completeness, $\int \di\mu(h)E_h = I_{AB}$. To show this, observe that
\begin{align}
\int \di\mu&(h)E_h=D_{s_A}D_{s_B}\int\di\mu(g)\di\mu(h)\kb gg_A\ot \kb{gh}{gh}_B\nonumber \\
&=\Bigl(D_{s_A}\int\di\mu(g)\kb gg_A\Bigr)\ot \Bigl(D_{s_B} \int\di\mu(h) \kb{h}{h}_B\Bigr) \nonumber \\
&=I_{AB}\,,
\end{align}
with the second line obtained by measure invariance, and the last using the property of maximum likelihood states $\m G(|e\rangle_A \langle e|) = D_{s_A}^{-1} I_A$ and $\m G(|e\rangle_B \langle e|) = D_{s_B}^{-1} I_B$.

With each effect $E_h$, we can define a corresponding CP map $\m M_{AB}^h$ describing both the measurement and subsequent update map in terms of the projectors as
\begin{equation}
\M^h_{AB}(\rho_{AB})=D_{s_A}D_{s_B}\int\di\mu(g)\Pi^{g,h}_{AB} \;\rho_{AB}\;{\Pi^{g,h}_{AB}}^\dag\label{eq-M}.
\end{equation}
Note that this update map is chosen such that the measurement is repeatable.  As with the POVM, this map can be implemented without the use of a background frame.  We prove this fact by demonstrating that the map is group-invariant, which means that the measurement map \eqref{eq-M} is invariant under any global rotation $\m U_{AB}(f):=\m U_A(f)\otimes\m U_B(f)$, i.e.\;for any $f\in G$ we have that~\cite{BartlettWiseman03,GourSpekkens}
\begin{equation}
\m U_{AB}(f)\circ\m M_{AB}^h\circ\m U_{AB}^\dag(f)=\m M_{AB}^h\,,
\end{equation}
To show this, observe using~\eqref{eq-M} that
\begin{multline}
\m U_{AB}(f)\circ\M^h_{AB}\circ\m U_{AB}^\dag(f)[\sigma]=D_{s_A}D_{s_B}\times\\\int\di\mu(g)U_{AB}(f)\Pi^{g,h}_{AB} \,U_{AB}^\dag(f)\sigma U_{AB}(f)\,{\Pi^{g,h}_{AB}}^\dag U_{AB}^\dag(f)\,.
\end{multline}
Focusing on just the projectors, from \eqref{eq-proj} we have that
\begin{equation}
U_{AB}(f)\Pi_{AB}^{g,h}U_{AB}^\dag(f)=\proj{fg}_A\otimes\proj{fgh}_B = \Pi_{AB}^{fg,h}.
\end{equation}
The group invariance of the integration measure in \eqref{eq-M} allows us to redefine $fg\to g$, thereby recovering the original map.

Note that, for a nonabelian symmetry group, this $G$-invariance of the map constrains the construction of the projectors.  If we had instead defined the projectors as $\kb{g}{g}_A\ot\kb{hg}{hg}_B$, the resulting map would not be $G$-invariant except in the special case of $h$ satisfying $hg=gh$ for all $g\in G$ (i.e.\;$h$ in the centre of the group).

\subsubsection{Using the relational measurement to change quantum reference frames\label{sec-specialcases}\label{sec-mainmeasurement}}

We have constructed a quantum operation \eqref{eq-genCQRF} to determine the relative orientation between two quantum reference frames.  In this section we will apply this operation to the problem of changing quantum reference frames. We will show that this measurement can be used to construct a quantum operation that  takes system $\rho_S$ encoded with respect to the quantum reference frame $\rho_A$ on $A$, i.e., the encoded state $\m E_{\rho_A}(\rho_S)$, and transforms it to a state $\rho_S'$ encoded with respect to a new reference frame $\rho_B$ on $B$ as the encoded state $\m E_{\rho_B}(\rho_S')$. We will quantify the performance of this task, i.e., how well the encoding of $\rho_S$ is preserved, in \secref{sec-Moptimal}.

Consider the action of the measurement map $\m M_{AB}^h$ given by \Eeqref{eq-M} on generic $G$-invariant states $\sigma_{SAB}$ on systems $S,A,B$ (defined to act as the identity map on $S$).  Because it is $G$-covariant, the map will produce a $G$-invariant state on $SAB$. For the purposes of the change of quantum reference frame procedure we want a map from $SAB$ to $SB$, as we want to discard the $A$ reference frame following the measurement. This is done by applying a partial trace over $A$ to the post-measurement state. The result is a final (unnormalised) $G$-invariant state on systems $S$ and $B$ with correlation between the subsystems. The unnormalised final state corresponding to measurement outcome $h$ is
\begin{multline}
\Tr_A[\M_{AB}^h(\sigma_{S AB})]=D_{s_A}D_{s_B}\times\\\int\di\mu(g)\bigl[\bra g_A\ot\bra{gh}_B\bigr]\;\sigma_{S AB}\bigl[\ket {g}_A\ot\ket{gh}_B\bigr]\otimes\proj{gh}_B\label{eq-genCQRF}.
\end{multline}
The measurement outcome $h$ is a continuous parameter, so we have a probability density function for outcomes $h$ for the measurement of a state $\sigma_{SAB}$ given by
\begin{equation}
P(h)=\Tr[E_h\sigma_{SAB}]= \Tr[\m M^h_{AB}(\sigma_{SAB})]\,.\label{eq-P}
\end{equation}
The probability density function normalises by $\int P(h)\di\mu(h)=1$ when using the group-invariant Haar measure $\di\mu(g)$.

Consider a relational encoding of a quantum state $\rho_S$ using a quantum reference frame $\ket{\psi(a)}_A$, a pure state with a well-defined orientation $a\in G$. (As we define the procedure to act on encoded states, the parameter $a$ describing the orientation of $A$ relative to a background has no operational significance. However, for the purposes of clarity, we leave this parameter $a$ in the derivation as it takes the role of $x_A$ from the example of \secref{sec-introchangingRFs}.) We are particularly interested in the case where this is a maximum likelihood state $\ket{\psi(a)}_A = U(a)\ket e_A$ for $a\in G$, although our map can be defined for a general quantum reference frame state.
No other reference frame is implicated, so we describe the joint $SA$ system by the $G$-twirled state $\m E_{\ket{\psi(a)}_A}(\rho_S)=\m G_{S A}(\rho_S\otimes\proj{\psi(a)}_A)$. We introduce a second reference frame $\rho_B$ which is non-implicated, i.e., uncorrelated with the other two quantum systems, described by the state $\m G_B(\rho_B)$. The full initial state on all components (the system $S$ and both quantum reference frames $A$ and $B$) is then
\begin{equation}
\sigma_{S AB} = \m G_{S A}(\rho_S\otimes\proj{\psi( a)}_A)\otimes\m G_B(\rho_B).\label{eq-uncorinitstate}
\end{equation}

We apply the operation $\sigma_{SAB} \rightarrow \Tr_A[\m M_{AB}^h(\sigma_{SAB})]$ given by \eqref{eq-genCQRF} to this state $\sigma_{SAB}$. This state is group-invariant, satisfying $\m G_{SAB}(\sigma_{SAB})=\sigma_{SAB}$, and $\m M_{AB}^h$ is $G$-covariant, so we can commute the $G$-twirl with the operation, allowing us to write the final state on $SB$ as
\begin{widetext}
\begin{align}
\Tr_A[\M_{AB}^h(\sigma_{SAB})]
=&\m G_{SB}\Bigl[\rho_S\otimes\bigl(D_{s_A}D_{s_B}\int\di\mu(g)\!\abs{\bk g{\psi( a)}_A}^2\!\bra{gh}\!\m G_B(\rho_B)\!\ket{gh}\;\proj{gh}_B\bigr)\Bigr] \nonumber \\
=&\m G_{SB}\Bigl[\rho_S\otimes\bigl(D_{s_A}\int\di\mu(g)\, \abs{\bk g{\psi( a)}_A}^2\;\proj{gh}_B\bigr)\Bigr]\label{eq-finalstate}
\end{align}
\end{widetext}
where the second line follows from the simplification $\bra{gh}\m G_B(\rho_B)\ket{gh}=\Tr[\m G(\proj{e}) \rho_B]=D_{s_B}^{-1}$, arising from properties of the $G$-twirl and maximum likelihood states $\ket e$. As $B$ was initially in the $G$-invariant state $\m G_{B} (\rho_B)$, all measurement outcomes $h$ are equally likely, and the result of the measurement is to initialise a reference frame state $\ket{gh}_B$ on $B$ with a well-defined orientation with respect to $A$.  Solving \eqref{eq-P} using \eqref{eq-finalstate}, we have that $P(h)=1$ for input states of the form \eqref{eq-uncorinitstate}. We can therefore associate $\Tr[\m M^h_{AB}(\sigma_{SAB})]$  with a trace one normalised state.

To continue simplifying \eqref{eq-finalstate}, the $G$-twirl $\m G_{SB}$ allows us to move the action of $g$ onto the state of the system $S$, as
\begin{align}
\Tr_A[\M_{AB}^h(\sigma_{SAB})]&=\notag\\
\m G_{SB}\Bigl[\notag D_{s_A}\int\di\mu(g)&\, \abs{\bk g{\psi( a)}_A}^2\;\m U_S(g^{-1}) [\rho_S]\bigr) \otimes\proj{h}_B \Bigr]  \nonumber\\
& =\m G_{SB}\left(\rho_S'\otimes\kb hh_B\right) \,,\label{eq-Trenc}
\end{align}
where we have defined
\begin{equation}
  \rho_S' = D_{s_A}\int\di\mu(g)\, \abs{\bk g{\psi( a)}_A}^2\;\m U_S(g^{-1}) [\rho_S] \,.
\end{equation}
With $\ket{\psi(a)}$ covariant, we have that $\bk g{\psi(a)}_A=\bk{a^{-1}g}{\psi(e)}_A$. Redefining $g\to ag$, we have that the new encoded system state $\rho_S'$ is related to the original system state $\rho_S$ by the composition of maps
\begin{equation}
\rho_S'=\m F_S^{(A)}\circ\m U_S(a^{-1})[\rho_S]\label{eq-rhoFUrho}
\end{equation}
where the form of the CP map $\m F_S^{(A)}$ is
\begin{equation}
\m F_S^{(A)}:=D_{s_A}\int\di\mu(g)\abs{\bk{g}{\psi(e)}_A}^2\m U_S(g^{-1}).\label{eq-genCQRFG}
\end{equation}
Note that the map $\m F_S^{(A)}$ is trace-preserving since $\int\di\mu(g)\abs{\bk g{\psi(e)}_A}^2=D_{s_A}^{-1}$. From \Eeqref{eq-Trenc} we now see explicitly that the result of the relational measurement, followed by tracing out of $A$, results in a final state that takes the form of $\rho'_S$ encoded with respect to a quantum reference frame on ${\ket h}_B$ on $B$, i.e., a state of the form
\begin{equation}
\Tr_A[\m M_{AB}^h(\sigma_{SAB})]= \m E_{\ket h_B}(\rho_S')
,\label{eq-mainmeas}
\end{equation}
which depends explicitly on the measurement outcome $h$. The CP map $\m F_S^{(A)}$ that takes $\rho_S$ to $\rho_S'$ is a convex mixture of unitary maps determined by the overlap of a maximum likelihood state with the reference frame state on $A$. Therefore, in general, this map results in decoherence of $\rho_S$ due to the uncertainty in orientation of the reference frame state $\ket{\psi(e)}_A$.

\subsection{Decoherence and performance of the change of quantum reference frame procedure\label{sec-dec2}}

\begin{figure}
\[
\xymatrix{
& \rho_S' \ar[ddr]^{\hbox{Encoding $\m E_{\ket{h}_B}$}}&\\
&{\m E_{\ket h_B}\!\circ\!\m R}&\\
\m E_{\ket{\psi(a)}_A}(\rho_S) \ar[uur]^{\hbox{Recovery $\m R$}} \ar[rr]_{\hbox{Change of frame}}^{\Tr_A[\m M_{AB}^h(*\otimes\sigma_B)]} &&  \m E_{\ket h_B}(\rho_S')
}
\]
\caption[Re-encoding with or without using a background frame]{The procedure \eqref{eq-genCQRF} on states of the form given by \eqref{eq-uncorinitstate} is the same map as $\m E_{\ket h_B}\!\circ\m R$ on an encoding $\m E_{\ket{\psi(a)}_A}(\rho_S)$. The change of quantum reference frame map results in description of the state relative to a  different reference frame without measuring against a background classical frame, whereas the re-encoding map $\m E_{\ket h_B}\!\circ\m R$ changes a reference frame using a background reference frame for the intermediate state $\rho_S'$. \label{fig-CRE}}
\end{figure}
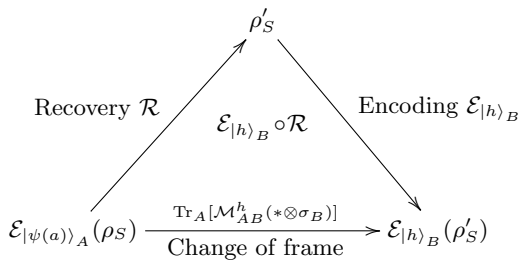

We will now characterise the decoherence of the system $\rho_S$ due to the change of quantum reference frame procedure. In particular, we will show that the decoherence, which is described by the CP map $\m F_S^{(A)}$, is equivalent to the decoherence associated with the `recovery' operation defined in \secref{sec-introdequant}.

For a system $S$ in a relational encoding with a quantum reference frame $A$, the recovery map $\m R$ of \eqref{eq-rec} can be implemented by measuring the orientation $g$ of the quantum reference frame $A$ relative to a background classical frame, discarding the quantum reference frame, and rotating $\rho_S$ by $g^{-1}$. This recovery map applied to a relational encoding of a system $\m E_{\ket{\psi(a)}_A}(\rho_S)$ leads to a noise map on $\rho_S$:
\begin{equation}
\m R\circ\m E_{\ket{\psi(a)}_A}(\rho_S)=D_{s_A}\int\di\mu(g)\abs{\bk g{\psi(a)}_A}^2\m U_S(g^{-1})[\rho_S].\label{eq-REinterp}
\end{equation}
This map is identical to the decoherence map $\m F_S^{(A)}\circ\m U_S(a^{-1})$ in \Eeqref{eq-rhoFUrho}. We then have the equivalence of maps
\begin{equation}
\m R\circ\m E_{\ket{\psi(a)}_A}\equiv\m F_S^{(A)}\circ\m U_S(a^{-1})\label{eq-REFU}.
\end{equation}
As depicted in \figref{fig-CRE}, the transformation $\m E_{\ket{\psi(a)}_A}(\rho_S)\mapsto\m E_{\ket h_B}(\rho_S')$ achieved by the change of quantum reference frame procedure can also be achieved by composing the recovery and encoding maps $\m E_{\ket h_B}\circ\m R$. We can therefore write the final state of the change of quantum reference frame procedure as
\begin{equation}
\m E_{\ket h_B}(\rho_S')=\m E_{\ket h_B}\left[(\m R\circ\m  E_{\ket{\psi(a)}_A})[\rho_S]\right].\label{eq-finstateRE}
\end{equation}
We have shown that the change of quantum reference frame procedure results in a encoded state $\m E_{\ket h_B}(\rho_S')$ where $\rho_S'$ is related to the initial state $\rho_S$ (in the encoded state $\m E_{\ket{\psi(a)}_A}(\rho_S)$) by the map $\m R\circ\m E_{\ket{\psi(a)}_A}$. Let us now view the change of quantum reference frame procedure in terms of the operational task set out at the start of \secref{sec-genresults}.

\subsubsection{Classical limits\label{sec-classicallimits}}

Recall the notation introduced in \secref{sec-groupestates}, with the states of quantum references frames $\ket{s;\psi(g)}$ parameterised by a size parameter $s$, and for which $s \to \infty$ describes the classical limit.
Given that decoherence occurs in the change of quantum reference frame procedure due to the uncertainties in orientation of each reference frame state $\ket{s;\psi(g)}$, we want to identify the conditions under which there is no decoherence. For a class of quantum reference frame states $\ket{s;\psi(g)}$ that possess a well-defined classical limit $s\to\infty$ in which there is no uncertainty in orientation (for example, the maximum likelihood states $\ket{s;g}$), we demonstrate that the change of quantum reference frame procedure has an appropriate classical limit.

To reproduce a classical change of reference frame map, for size parameters $s_A,s_B\to\infty$ we should have the initial relational encoding $\lim_{s_A \to \infty} \m E_{\ket{s_A;\psi(a)}_A}(\rho_S)$ map to the final relational encoding $\lim_{s_B \to \infty} \m E_{\ket{s_B;ah}_B}(\rho_S)$ with no change to the encoded state $\rho_S$, i.e., no decoherence.  We now show that this is the case. As $s_A\to\infty$, the overlap of $\ket{s_A;\psi(e)}$ with other orientations in the group becomes zero, i.e., we have
\begin{equation}
   \lim_{s_A \to \infty} D_{s_A}\abs{\bk {s_A;g}{s_A;\psi(e)}_A}^2 = \delta(g)\,,
\end{equation}
where $\delta$ is the Dirac delta function on the group. The decoherence map \eqref{eq-genCQRFG} then becomes the identity map, $\lim_{s_A \to \infty} \m F_S^{(A)}=\m I_S$. The final state is then $\sigma_{S B}^h =\m E_{\ket{s_B;ah}_B}[\rho_S]$ where the size of the reference frame $B$ is determined by the size $s_B$ of the initial $B$ reference frame. This reproduces the required classical limit.

We can also consider the single limits where only one of the reference frames $A$ or  $B$ is taken to be classical. For the case where we take the classical limit of $A$, the change of quantum reference frame procedure is simply the encoding map $\m E_{\ket{ah}_B}(\rho_S)$. In the alternate case where $A$ remains finite but $B$ becomes classical, the operation is simply the recovery map. This classical limit corresponds to recovery from a quantum frame $A$ into a classical frame $B$ with associated decoherence $\m F_S^{(A)}$.

\subsubsection{Performance of the procedure for changing quantum reference frames\label{sec-Moptimal}}

In \secref{sec-genresults} we defined the operational task for changing quantum reference frames as follows: an observer possesses a quantum system $S$ and implicated quantum reference frame $A$ as the initial encoded state  $\m E_{\rho_A}(\rho_S)$. This observer wishes to use a second, initially non-implicated, quantum reference frame $B$ as a quantum reference frame for the system $S$, and to discard the initial reference frame $A$, resulting in a final encoded state $\m E_{\rho_B}(\rho_S')$.

We can now quantify the performance of our change of reference frame procedure defined in \secref{sec-genresults} for this operational task. We can view any such procedure as a map $\m O:\rho_S\mapsto\rho_S'$ and quantify its performance according to some figure of merit such as its process fidelity. We have shown in \secref{sec-dec2} that this map for our change of reference frame procedure is the decoherence map $\m F_S^{(A)}\circ \m U_S(a^{-1})=\m R\circ\m E_{\ket{\psi(a)}_A}$. As $\m U_S(a^{-1})$ is unitary, any loss in fidelity is due to the map $\m F_S^{(A)}$.

To determine whether our map is optimal relative to some figure of merit, i.e.\;whether the map induces the least amount of decoherence for the task outlined above, is a difficult problem in general. However, we note that in the classical limits defined in \secref{sec-classicallimits}, the associated decoherence is that corresponding to either $\m E_{\ket{ah}_B}$ or $\m R$ for a given reference frame, and these maps are shown to be optimal and near-optimal in \refcite{BarnumKrill02} (see \secref{sec-REOptimal}). Therefore our change of reference frame procedure is near optimal in these classical limits.

\subsection{Consequences and interpretation of decoherence\label{sec-interpretation}}

In the previous sections, we have developed the mathematical tools to describe the change of a quantum reference frame. Before investigating two examples in Secs.~\ref{sec-phaseex} and \ref{sec-exampleSU2}, it is worthwhile to consider at this stage some of the conceptual consequences of the procedure.

As we identified in \secref{sec-dec2}, following the change of quantum reference frame procedure the system in the final encoded state appears to be affected by a form of decoherence.  This decoherence is absent in the classical limit.  In this section we will investigate the properties of the decoherence, the necessity of its existence in a change of reference frame procedure, and consider the consequences for the relativity principle for quantum reference frames, suggesting a connection to a type of \textit{intrinsic decoherence}.

\subsubsection{Properties of the decoherence from changing quantum frames}

First, we pose some questions regarding the properties of the decoherence in the procedure:  Is decoherence necessary when changing a quantum reference frame? Could the decoherence be reduced by changing to a better (more precise) reference frame?

In \secref{sec-dec2} we determined that the decoherence due to changing quantum reference frames with the procedure is associated with the map $\m F_S^{(A)}$, a convex mixture of unitary maps determined by the overlap $\abs{\bk g{\psi(e)}_A}^2$ of a maximum likelihood state with the reference frame state on $A$. The states in this overlap are generally not orthogonal unless the reference frame $A$ approaches infinite size,

and so the change of reference frame procedure will cause decoherence. Additionally, unless reference frame $B$ is of infinite size, there is also decoherence associated with the encoding with respect to the quantum frame $B$. The net decoherence on the system is the composition of these two sources. As a consequence, changing from a less precise frame $A$ to a more precise frame $B$ nonetheless still results in a net increase in the decoherence to the system.

\subsubsection{Interpreting the decoherence as an intrinsic decoherence}

Now that we have identified the decoherence as being fundamental to change of quantum reference frames, there is still the question of how to view the decoherence arising as a result of a change of quantum reference frame in the context of the relativity principle. To this end, we will interpret the decoherence in terms of an intrinsic decoherence.  Intrinsic decoherence is decoherence to a quantum state that occurs without interaction with an environment \cite{Stamp12}.
It has been proposed to occur as a result of fluctuations in the spacetime metric or other aspects of background spacetime due to quantum effects of the spacetime in theories of quantum gravity \cite{Milburn91,Power00,Gambini04,KokYurtsever,GirelliPoulin08}. By internalising parameters into a quantum model, quantum reference frames provide a way to model the effects of a background spacetime. A connection between deformed symmetries of semiclassical gravity and quantum reference frames was demonstrated in \refcite{GirelliPoulin07}. Most closely related to quantum reference frame measurement is by \citet{Milburn03} in which intrinsic decoherence arises when a quantum state is translated in position by an operator whose parameters are not precisely known, due to the quantisation and uncertainty of the background time parameter.

We will interpret the decoherence of the change of quantum reference frame procedure \Ssref{sec-introCoRF} within the spacetime intrinsic decoherence framework introduced above. The change of quantum reference frame procedure is a complete, closed description of the decoherence that occurs to a system $\rho_S$ due to changing between two quantum reference frames. The corresponding description of a change of reference frame when the two reference frames are treated as background frames, so that only the system $\rho_S$ remains quantum mechanical, is that the system experiences a noise map $\m F_S^{(A)}$ as an isolated quantum system; i.e.\;in this description the quantum system experiences intrinsic decoherence. Viewed from the other perspective, we have that the change of quantum reference frame map is the self-contained description of this intrinsic decoherence once the quantum nature of the reference frames is included. As such it is an operational derivation of a process that leads to intrinsic decoherence. Note that this particular model consists of an abrupt measurement rather than dynamics or continuous time evolution.

\section{Example: Phase reference \label{sec-phaseex}\label{sec-genAbelgroup}}

In this section, we explore a simple example illustrating the details of the change of quantum reference frame procedure for reference frames associated with an Abelian group.  Specifically, we consider a phase reference, whose orientation corresponds to an element of $U(1)$.  We will pay particular attention to the explicit forms and interpretation of the final state described in \secref{sec-mainmeasurement} which results from a change of quantum reference frame procedure.

The example will be structured as follows. First we will describe the reference frame states we will use and how these allow storage of quantum information in relational degrees of freedom. We will then review the change of quantum reference frame procedure in this Abelian case. We then explicitly calculate the decoherence for the cases where the reference frame is described by a phase eigenstate or coherent state. There will be some comparison of the decoherence for these choices. We will use these results to verify the classical limits of the procedure as described in \secref{sec-classicallimits}. Finally, we comment on the similarity of the relational reference frame measurement to balanced homodyne detection, with details in \appref{app-BHD}.

\subsection{The representation of $U(1)$ on harmonic oscillators}

We first present the structure of the representation of $U(1)$ on a collection of harmonic oscillators, and how we might encode information in a relational way.  For an Abelian group, group multiplication becomes addition and the identity $e$ can be written as $0$. The unitary group $U(1)$ can be considered as the group of phases $\theta$ with the group multiplication being addition modulo $2\pi$. We will, however, retain the generic group element notation $g,a,h \in U(1)$ for familiarity with the general formalism. The charge sectors of the representation are subspaces of total photon number. The unitary representation of $U(1)$ on a single mode state is $U(g)=\Ee^{\ii \hat ng}$ where $\hat n$ is the number operator $\hat n\ket k_\text{Fock}=k\ket k_\text{Fock}$. The $U(1)$ Haar integration measure is $\di\mu(g)=\di g/2\pi$. Therefore, for a single mode harmonic oscillator, the $G$-twirl of a state $\sum_{k=0}^\infty a_k\ket k_\text{Fock}$ is
\begin{align}
\m G\Bigl(\sum_{k,l=0}^\infty a_ka_l^*&\kb kl_\text{Fock}\Bigr)\notag\\
&=\int_0^{2\pi}\frac{\di g}{2\pi}\Ee^{\ii(k-l)g}\sum_{k,l=0}^\infty a_ka_l^*\kb kl_\text{Fock}\notag\\
&=\sum_{k=0}^\infty \abs{a_k}^2\kb kk_\text{Fock}
\end{align}
with the integral giving the constraint $k=l$. The phase information in a single mode state is thus completely decohered. However, if we introduce a second mode, i.e., a second oscillator with distinguishable frequency, we can form the two-mode pure state $\ket{\psi_{SA}}=\sum_{k,l}a_{k,l}\ket{kl}_\text{Fock}$. Written in terms of total photons $2n=k+l$ and difference $2j=k-l$, where $n$ can take any non-negative half-integer value and where $j=-n,-n +1, \dots, n$ \cite{TycSanders}, this becomes $\sum_{n,j}a_{n+j,n-j}\ket{n+j,n-j}_\text{Fock}$. The $G$-twirl on this state is
\begin{widetext}
\begin{align}
\m G_{SA}(\proj\psi_{SA})=&\sum_{n,j}\sum_{m,k}\int \frac{\di g}{2\pi}\Ee^{\ii(2n-2m) g}\Big(a_{n+j,n-j}\ket{n+j,n-j}_\text{Fock}\Big) \Big(a^*_{m+k,m-k}\bra{m+k,m-k}_\text{Fock}\Big) \nonumber \\
=&\sum_n\Big(\sum_{j,k}a_{n+j,n-j}a_{n+k,n-k}^*\kb{n+j,n-j}{n+k,n-k}_\text{Fock}\Big).
\end{align}
\end{widetext}
Phase coherence remains within subspaces of total photon number eigenstates, producing a total state that is a mixture over total photon number $2n$ of pure eigenstates $\sum_{j}a_{n+j,n-j}\ket{n+j,n-j}_\text{Fock}$ of total photon number $2n$. With judicious choices of a reference frame state on $A$, a state on $S$ can be relationally encoded into the subspaces of total photon number~\cite{SBRK,BRSDialogue}.

\subsection{Reference frames for $U(1)$}

We define our two reference frames $A$ and $B$ to be single mode harmonic oscillators in group-covariant states $\ket{\psi(gh)}=U(g)\ket{\psi(h)}$. The particular examples we will study are the $U(1)$ maximum likelihood states, and $U(1)$ coherent states, both of which have well-defined size parameters. In this example, the recovery map using the maximum likelihood states is optimal \cite{BRST}, as defined in \secref{sec-REOptimal}, and so provides the optimal change of quantum reference frame procedure as defined in \secref{sec-Moptimal}.

\subsubsection{Reference frame $A$ in phase eigenstate\label{sec-BHD}}

The maximum likelihood states (introduced in \secref{sec-groupestates}) for a representation of the $U(1)$ group on a single mode Fock space truncated in maximum photon number $s$ are the bounded-size phase eigenstates with photon number cutoff $s$  \cite{PeggBarnett97}. The phase eigenstate with phase $g$ and size parameter $s$ is given by
\begin{equation}
\ket{s;g}:=N_s^{-\half}\sum_{k=0}^s\Ee^{\ii k g}\ket k_\text{Fock}\label{eq-phaseeigenstate}
\end{equation}
where $\ket k_\text{Fock}$ is the Fock state with $k$ excitations, and the state normalisation is $N_s=(s+1)=D_s$, the dimension of the Hilbert space.

In addition, as these states satisfy $\m G (\proj {s;g}) = (s+1)^{-1}I_s$, they will also be used to form the projectors \eqref{eq-proj} for measurement.

We now consider the change of quantum reference frame procedure for the $U(1)$ group, using phase eigenstates both for our initial reference frame on $A$ as well as forming the relational measurement.  In this procedure, an initial state $\sigma_{S AB}=\m G_{S A}(\rho_S\otimes\kb{\psi(a)}{\psi(a)}_A)\otimes \m G_B(\rho_B)$ is transformed to the final state for outcome $h$ on $SB$ given by
\begin{equation}
\sigma_{SB}^h=\Tr_A\bigl[\m M_{AB}^h(\sigma_{SAB})\bigr]=\m E_{\ket{a+h}_B}\bigl(\m F_S^{(A)}[\rho_S]\bigr)\,,
\end{equation}
where we have commuted the rotations $a$, $h$, and the map $\m F_S^{(A)}$ due to $U(1)$ being Abelian. For the state of reference frame $A$ prepared in the bounded-size phase eigenstate $\rho_A=\proj{s_A;a}$ with cutoff $s_A$, the overlap between two phase eigenstates with cutoffs $s$ gives \cite{PeggBarnett89}
\begin{align}
\abs{\bk{s;g}{s;h}}^2=&D_s^{-2}\sum_{k=-s}^s(s+1-\abs k) \Ee^{\ii k(h-g)}\label{eq-U1overlap}\\
=&\frac{1}{(s+1)^2}\frac{1-\cos[(s+1)(h-g)]}{1-\cos[h-g]}\notag.
\end{align}
The measurement of relative orientation $h$ is constructed from a family of projectors \eqref{eq-proj} on the $A$ and $B$ Hilbert spaces. In this example the projectors will be constructed in terms of $U(1)$ maximum likelihood states with size cutoffs $s_A,s_B$. Due to the equally weighted superposition of number states of the phase states in the projectors, a measurement constructed from such a family of projectors resolves the identity on the space of the reference frames $A$ and $B$. The sizes of the projectors is set to be equal to the cutoff of the reference frame states, $s_A$ and $s_B$. The decoherence map \eqref{eq-genCQRF} then takes the form
\begin{align}
\m F_S^{(A)}=&D_{s_A}\int\frac{\di g}{2\pi}\abs{\bk{s_A;g}{s_A;0}_A}^2 \m U_S(-g)\label{eq-abelF}\\
=&\frac{1}{(s+1)}\int\frac{\di g}{2\pi}\frac{1-\cos[(s_A+1)g]}{1-\cos g}\m U_S(-g).\label{eq-mixedU1PEf}
\end{align}

The distribution of unitaries in $g\in G$ is graphed in \figref{fig-mixedU1f} for average photon number $\ex n_A=s_A/2=4$ and $8$. The function is symmetric about $g=0$, at which it is peaked.

We note that the relational measurement has many similarities to \emph{balanced homodyne detection}:  a measurement technique from quantum optics.  We explore this relationship in \appref{app-BHD}.

\subsubsection{Reference frame $A$ in coherent state\label{sec-U1cohstate}}

We also consider reference frame $A$ given by a coherent state
\begin{equation}
\ket{s_A;g}_{\rm CS}=\Ee^{-s_A^2/2}\sum_{k=0}^\infty \frac{s_A^k\Ee^{\ii kg}}{\sqrt{k!}}\ket k_\text{Fock}.\label{eq-U1CS}
\end{equation}
The coherent state has a well-defined phase $g$ (i.e.\;orientation in $U(1)$) and transforms covariantly under the group: $U(g)\ket{s_A;0}_{\rm CS}=\ket{s_A;g}_{\rm CS}$. It has a size $s_A$ characterised by the square root of the mean photon number, $s_A=\sqrt{\ex n}$. The $G$-twirl of this state gives a Poisson distribution in photon number, with no phase coherence.

Although coherent states are suitable as quantum reference frames, there are challenges to constructing relational measurements using projectors onto these states because $\m G(\proj{s;g})$ is not proportional to the identity.  We therefore restrict to the relational measurement constructed out of phase eigenstate projectors.

Coherent states have non-zero support on all photon numbers $n\to\infty$, so we will use an infinite limit for the size $s_A$ of the projectors on $A$ for this example. The POVM will resolve the identity on the full infinite dimensional Fock space. We will need to keep in mind that the initial $B$ state may also have $s_B\to\infty$, for example, if it is a mixture of coherent states, in which case the projectors on $B$ and consequently the post-measurement state on $B$ will have infinite size.

\begin{figure}[h]
\centering
\includegraphics[height=5.8cm]{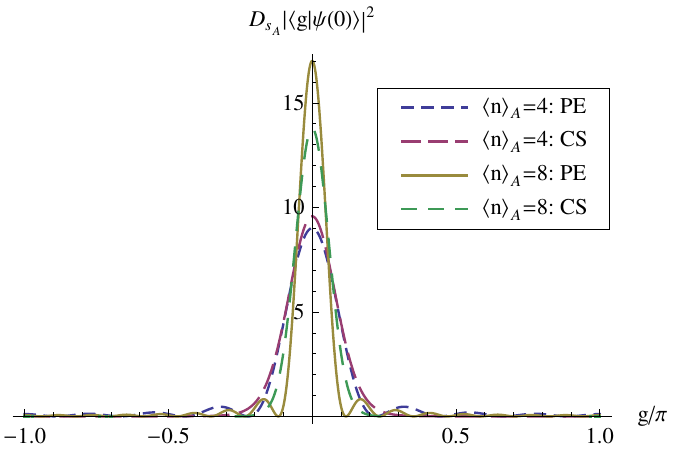}
\caption[State overlaps for $U(1)$ frames (colour)]{(Color online) Plotted are the state overlaps $D_{s_A}\abs{\bk g{\psi(0)}}^2$ for reference frame $A$ in a $U(1)$ phase eigenstate (\eqref{eq-U1overlap}, `PE', solid lines) and coherent state (\eqref{eq-U1CSoverlap}, `CS', broken lines), for choices of average photon number $\ex n_A$. This indicates the distribution of unitaries in the decoherence maps $\m F_S^{(A)}$. For small average photon number the decoherence for the coherent states has a narrower peak than the phase eigenstate, but the phase eigenstate becomes more narrowly peaked by $\ex n_A=5$. For calculations the summations for the coherent state overlap were truncated at the 21st terms, accounting for 99.99\% of the support. \label{fig-mixedU1f}}
\end{figure}

The overlap of a coherent state with a phase eigenstate used in the projectors is
\begin{equation}
\bk{s;g}{t;h}_{\rm CS}=D_s^{-\half}\Ee^{-t^2/2}\sum_{k=0}^s\frac{t^k\Ee^{\ii k(h-g)}}{\sqrt{k!}}
\label{eq-U1CSoverlap}
\end{equation}
where we take the support of the projectors $s\to\infty$.  (Because the POVM has normalisation factors $D_s$, this limit will still result in a well-defined projector.)  The decoherence map \eqref{eq-genCQRFG} is then
\begin{equation}
\m F_S^{(A)}=\int\di\mu(g) \Ee^{-t^2/2}\sum_{k=0}^s\frac{t^k\Ee^{-\ii kg}}{\sqrt{k!}}\m U_S(-g)\label{eq-mixedU1CSf}
 \end{equation}
The distribution of the unitaries in this decoherence map is plotted in \figref{fig-mixedU1f} for choices of $s_A$, and compared with the corresponding phase eigenstate distribution \eqref{eq-mixedU1PEf} for the same average photon number.

\subsection{Classical limits}

We briefly examine and interpret the results of the change of reference frame procedure for the classical limits of reference frames $A$ and $B$, i.e., when one or both of the size parameters $s_A,s_B$ are taken to infinity.

We will examine the $B$ classical limit first. The decoherence map \eqref{eq-mixedU1PEf} is not dependent on the $B$ frame, so it does not change in the $s_B\to\infty$ limit. The final state then has decoherence due to the finite size of the $A$ reference frame. If $B$ is initially in a mixture of a finite size phase eigenstate, then the effect of $s_B\to\infty$ is merely to increase the size of the final reference frame $B$ to its classical limit. For the coherent state example, the final state on $B$ is already an infinite-cutoff phase eigenstate. The interpretation of this limit is a recovery from quantum reference frame to classical frame, with noise accumulated solely due to the encoding with reference frame $A$.

To compute the decoherence map in the limit $s_A\to\infty$ we want to show that the overlap functions \eqref{eq-U1overlap} and \eqref{eq-U1CSoverlap} approach perfect distinguishability. For the phase eigenstate, using \eqref{eq-U1overlap} we can show that in the limit $s_A\to\infty$ the term becomes a delta function
\begin{align}
\lim_{s_A \to \infty} D_{s_A}&\abs{\bk{s_A;g}{s_A;h}}^2\notag\\
&=\lim_{s_A \to \infty}\sum_{k=-s_A}^{s_A}\frac{s_A+1-\abs k}{s_A+1}\Ee^{\ii k(h-g)}\notag\\
&= \sum_{k=-\infty}^{\infty}\Ee^{\ii k(h-g)}=\delta(h-g)\,,
\label{eq-U1peoverlaplimit}
\end{align}
where the denominator is provided by the state normalisation $D_{s_A}=s_A+1$ \eqref{eq-phaseeigenstate} and $\delta$ is normalised in the Haar measure: $\int_0^{2\pi}\delta(g)\frac{\di g}{2\pi}=1$. For coherent states, rather than attempting to directly compute the limit of the overlap, there are existing results we can use: A phase operator can be defined in terms of the states $\ket\theta=\sum_{n=0}^\infty \Ee^{\ii n\theta}\ket n_\text{Fock}$ \cite{WallsMilburn}. These are the same operators that we use in the projectors, so characteristics of phase indicate characteristics of the overlap function \eqref{eq-U1CSoverlap}. Indeed the operator is used to define a phase distribution $P(\theta)=\abs{\bk\theta\psi}^2/2\pi$ for some state $\psi$. Particularly, for $\psi$ a large coherent state $\ket{s,\phi}$, the mean of the phase distribution is $\ex{\theta}=\phi$ and the standard deviation is $\Delta\theta=\frac1{2s}$ \cite{WallsMilburn}. Then, as $s\to\infty$, the phase uncertainty becomes $\Delta\theta\to0$. Therefore we have $\lim_{s\to\infty}\abs{\bk{s,\phi}{\theta}}^2$ is non-zero only for $\theta=\phi$, for which the value is 1.

\section{Example: Cartesian and Direction frames \label{sec-exampleSU2}}

In this section we will consider the change of reference frame procedure for reference frames based on a nonabelian group, $SU(2)$, which describes the orientations of a Cartesian reference frame for three dimensions. We also consider a `direction indicator' state for three dimensions, which, due to rotational invariance around the single indicated direction, is associated with the coset space $SU(2)/U(1)$. We use $SU(2)$ rather than $SO(3)$ so that we can use spin representations.

The representation of $SU(2)$ decomposes a Hilbert space into a tensor sum of charge sectors of total spin $j$, where $j$ is a positive integer or half integer. In general, each of these is a reducible representation which can be further decomposed into a subsystem $\m M_j$ carrying an irreducible representation in a tensor product with a multiplicity subsystem $\m N_j$ which carries the trivial representation. The Hilbert space of a reference frame state would then decompose as $\Hi_A=\bigoplus_{j} \m M_A^{(j)}\otimes\m N_A^{(j)}$ \cite{BRST}.

\subsection{$SU(2)$ fiducial states (Cartesian frame)}

We define our reference frame systems using a Hilbert space $\m H_R=\bigoplus_{j} \m M_R^{(j)}\otimes\m N_R^{(j)}$, with the dimensions of the subsystems $\m M_R^{(j)}$ and $\m N_R^{(j)}$ chosen to be equal.  Such a space carries the regular representation of $SU(2)$, where each irrep $j$ appears with multiplicity equal to its dimension.  Following \cite{BRST}, we define a fiducial Cartesian reference frame state, with truncation parameter $s$, to be
\begin{equation}
\ket{s;e}:=D_s^{-\half}\sum_{j=0}^s\sqrt{2j+1}\sum_{m=-j}^j \ket{j,m}_\text{rot}\otimes\ket{\phi_{j,m}}\label{eq-SU2fidstate}
\end{equation}
which has support on integer spin $j$ charge sectors up to $j=s$. Here, $\ket{j,m}_\text{rot}$ is an eigenstate of $J_z$, and these for $m=-j,-j+1,\dots, j$ form a basis for $\m M_A^{(j)}$, denoted by $\ket{\cdot}_\text{rot}$. The states $\ket{\phi_{j,m}}$ form a basis for $\m N_A^{(j)}$. Together $\sum_{m=-j}^j\ket{j,m}_\text{rot}\otimes\ket{\phi_{j,m}}$ forms a state in the spin-$j$ charge sector which is maximally entangled between the irreducible representation and multiplicity subsystems. The state normalisation is the dimension of the vector space that $\ket{e_A}$ spans, and is given by
\begin{equation}
D_s=\sum_{j=0}^s(2j+1)^2=\frac13(2s+1)(2s+3)(s+1)=\binom{2s+3}3.
\end{equation}

For rotations of these states under $SU(2)$ we will use the polar parametrisation:
\begin{equation}
U(g)=U(\omega,\theta,\phi)=\Ee^{\ii\omega\mathbf n\cdot\mathbf J}
\end{equation}
with $\omega$ the rotation angle, $\mathbf n=(\sin\theta\cos\phi,\sin\theta\sin\phi,\cos\theta)$ the axis of rotation, $\frac\phi2,\theta,\omega\in[0,\pi)$, and with the Haar measure given by $\di\mu(g)=\sin^2\frac\omega2\sin\theta\ \di\phi\ \di\theta\ \di\omega/2\pi^2$.

\begin{figure}
\centering
 \includegraphics[height=4.4cm]{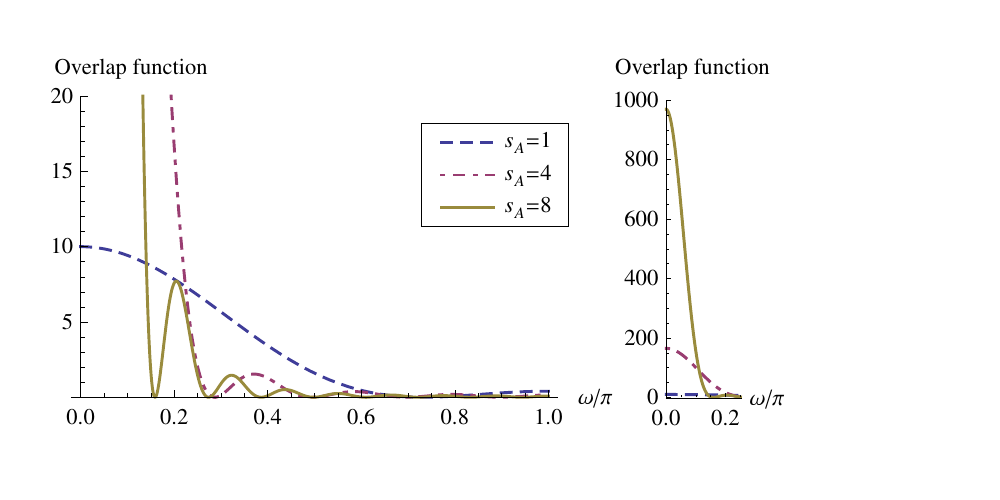}
\caption[State overlap in decoherence map for $SU(2)$ Cartesian fiducial states (color)]{(Color online) Plots of integrand of decoherence map \eqref{eq-mixedSU2GEf} for fiducial state Cartesian reference frames with $s_A=1,4,8$. The fiducial state overlap depends only on the rotation angle $\omega$ and not on the axis of the rotation $\mathbf n=(\sin\theta\cos\phi,\sin\theta\sin\phi,\cos\theta)$. The mixing is symmetric about the identity, identified by $\omega=0$. The first plot shows the details of the plot across $\omega$, and the second plot shows the full range for small $\omega$, indicating that overlap function becomes extremely peaked near $\omega=0$ even for small $s_A$, since the density of states approaches zero as $\omega\to0$.\label{fig-mixedSU2fPE}}
\end{figure}

For this example, we will use rotated fiducial states \eqref{eq-SU2fidstate} to form the measurement projectors \eqref{eq-proj}, with maximum $j$ cutoffs $s_A$ and $s_B$ for the projectors on $A$ and $B$, respectively. The overlap function of an unrotated fiducial state with an $SU(2)$-rotated state $\ket{s_A;g}=U(g)\ket{s_A;e}$ of the same size is
\begin{equation}
\bk{s_A;e}{s_A;g}
=D_{s_A}^{-1}\sum_{j=0}^{s_A}(2j+1) \chi^{(j)}(\omega,\theta,\phi)
\end{equation}
where $\chi^{(j)}(\omega,\theta,\phi)=\cos[(j+\half)\omega]/\cos(\omega/2)$ are the characters of $SU(2)$ \cite{BRST}. Using $\cos[(j+\half)\omega]/\cos(\omega/2)=\sum_{m=-j}^j\Ee^{\ii m\omega}$ and reordering summations (using $\sum_{j=m}^{s_A}(2j+1)=(s_A+1)^2-m^2$) we have
\begin{align}
\bk{s_A;e}{s_A;g}
=&D_{s_A}^{-1}\sum_{j=0}^{s_A}(2j+1) \sum_{m=-j}^j\Ee^{\ii m\omega}\notag\\
=&D_{s_A}^{-1}\sum_{m=-s_A}^{s_A}\Ee^{\ii m\omega}((1+s_A)^2-m^2).\label{eq-su2fiducialoverlap}
\end{align}
The decoherence map \eqref{eq-genCQRFG} is then
\begin{widetext}
\begin{equation}
\m F_S^{(A)}=\binom{2s_A+3}3^{-1}\int \frac{\di\omega\di\theta\di\phi}{2\pi^2}\sin^2\Bigl(\frac\omega2\Bigr)\sin\theta\; \Bigl(\sum_{m=-s_A}^{s_A}\Ee^{\ii m\omega}\bigl((1+s_A)^2-m^2\bigr)\Bigr)^2\m U_S(-\omega,\theta,\phi).\label{eq-mixedSU2GEf}
\end{equation}
\end{widetext}
The state overlap in this map is plotted in \figref{fig-mixedSU2fPE} for choices of $s_A$. The overlap function \eqref{eq-su2fiducialoverlap} is independent of the axis of rotation $\mathbf n$ of $g$, depending only on the rotation angle $\omega$.

\subsubsection{Classical limits of reference frame states}

Again, we can verify several classical limits:  the limit in which the $A$ reference frame becomes infinitely large; when the $B$ reference frame becomes infinitely large; and when both become infinitely large.

We can write the overlap function \eqref{eq-su2fiducialoverlap} as
\begin{equation}
\bk{s_A;e}{s_A;g}=D_{s_A}^{-1}\Bigl((1+s_A)^2+\frac{\di^2}{\di\omega^2}\Bigr)\sum_{k=-s_A}^{s_A} \Ee^{\ii k\omega}
\end{equation}
Then for the limit $s_A\to\infty$, we have, since $D_{s_A}\sim s_A^3$,
\begin{equation}
\lim_{s_A\to\infty}D_{s_A}^{-1}\Bigl((1+s_A)^2+\frac{\di^2}{\di\omega^2}\Bigr)\sum_{k=-s_A}^{s_A}\Ee^{\ii k\omega}= D_{s_A}^{-1}\frac2{\omega^2}\delta(\omega)\label{eq-SU2FSoverlaplimit}
\end{equation}
and in addition we enforce the normalisation condition $\bk{s_A;e}{s_A;e}=1$ for all $s_A$.

We can replace one inner product in the decoherence map \eqref{eq-mixedSU2GEf} with \eqref{eq-SU2FSoverlaplimit} to obtain $\frac2{\omega^2}\delta(\omega)\bk g e_A\simeq \frac2{\omega^2}\delta(\omega)$ in the integrand. Now integrating over $\omega$, the unitary is constrained to $\m U_S(0,\theta,\phi)=\m I$, and so the $\theta$ and $\phi$ integrals are trivial. The final state is then $\sigma_{S B}^h=\m G_{S B}\left[\rho_S\otimes\kb{ah}{ah}_B\right]$, mimicking an encoding from a classical frame to quantum frame.

The $s_B\to\infty$ limit results in an unchanged decoherence map, but an infinite reference frame on $B$ in the final state. This final state can be interpreted as a recovery \eqref{eq-rec} from a finite reference frame $A$ to infinite (`classical') reference frame $B$, where the mixing on the system $\rho_S$ is the decoherence due to the initial encoding with the imperfect $A$ reference frame.

The simultaneous infinite limit of $s_A$ and $s_B\to\infty$ then describes a change of classical reference frame operation.

\subsection{$SU(2)$ Coherent states:  A direction indicator}

As an illustrative example of the effect of the choice of fiducial state, we consider using an $SU(2)$ coherent state to define a direction reference frame.  Such a state indicates a direction on the two-sphere and has rotational symmetry (it is invariant up to global phase) about this direction. These $SU(2)$ coherent states reside within a single irreducible representation $\m M^{(j)}$ of $SU(2)$ and transform under $SU(2)$, but with a $U(1)$ invariance corresponding to the rotation about the direction in which the state is pointing. The set of possible orientations of a direction indicator therefore has the structure of a coset space $SU(2)/U(1)$, rather than a group. Consequently, the results in this example take different forms to the previous examples. Even in classical cases or limits of frames using this coset space, we will see dephasing operations on quantum systems due to the $U(1)$ rotational symmetry \cite{BRST}.

For this example we will use the Euler angle parametrisation of $SU(2)$ \cite{BRST}, as it allows us to easily separate the $J_z$ rotations under which the coherent states are invariant up to global phase
\begin{equation}
U(g)=U(\alpha,\beta,\gamma)=\Ee^{-\ii \alpha J_z}\Ee^{-\ii \beta J_y}\Ee^{-\ii \gamma J_z}\label{eq-SU2CSparam}
\end{equation}
with $\alpha,2\beta,\gamma\in[0,2\pi]$ and $\di\mu(g)=\di \alpha\,\sin \beta\, \di \beta\, \di \gamma/8\pi^2$. The coherent state corresponding to the identity orientation, on irreducible representation with total spin $j$, is defined and denoted as
$\ket{j;e}_{\rm CS}:=\ket{j,j}_\text{rot}$
and the $SU(2)$-rotated state $\ket{j;g}_{\rm CS}\equiv\ket{j;(\alpha,\beta,\gamma)}_{\rm CS}$ is \cite{Perelomov}
\begin{multline}
U(\alpha,\beta,\gamma)\ket{j,j}_\text{rot}=\\
\Ee^{-\ii\gamma j}\sum_{m=-j}^j{2j\choose j+m}^{\half}\cos^{j+m}\frac \beta2\sin^{j-m}\frac \beta2 \Ee^{-\ii\alpha m}\ket{j,m}_\text{rot}\\=:\ket{j;g}_{\rm CS}.
\label{eq-su2csrotate}
\end{multline}

The overlap of a rotated state with the identity coherent state is
\begin{align}
_{\rm CS}\bk{j;e}{l;g}_{\rm CS}&=\  _\text{rot}\!\bra{j,j}U(g)\ket{l,l}_\text{rot}\notag\\
&=\delta_{jl}\Ee^{-\ii(\alpha+\gamma)j} \cos^{2j}(\beta/2).\label{eq-su2csoverlap}
\end{align}
The $G$-twirl of \eqref{eq-su2csrotate} is $(2j+1)^{-1}I_{(2j+1)}$.

In this example the reference frame state size parameters $s_A$ and $s_B$ are given by the total spin $j$ of the coherent state. The measurement projectors \eqref{eq-proj} will consist of coherent states of the same sizes. The normalisation factors in the measurement \eqref{eq-M} are given by $D_{s}=2s+1$. The decoherence map is then
\begin{align}
\m F_S^{(A)}&=D_{s_A}\int\di\mu(g)\cos^{4s_A}\!\frac\beta2\,\m U_S(g^{-1})\nonumber\\
&=(2s_A+1)\times\notag\\
\int\!\frac{\di\alpha}{2\pi}&\frac{\di\gamma}{2\pi}\sin\!\beta\frac{\di\beta}2 \cos^{4s_A}\!\frac\beta2\m \,R_S^z(-\gamma)\circ\m R_S^y(-\beta)\circ\m R_S^z(-\alpha)\notag\\
=\m D_S\circ&\left[(2s_A+1)\int_0^\pi\sin\beta\frac{\di\beta}2 \cos^{4s_A}\!\frac\beta2\m R_S^y(-\beta)\right]\circ\m D_S.\label{eq-SU2CSF}
\end{align}
where $\m R_S^i(\theta)[\rho]:=\Ee^{-\ii \theta J_i}\rho\Ee^{\ii \theta J_i}$ is the superoperator for a unitary rotation of $\theta$ around the $i=y$ or $z$ axis and $\m D_S[\rho_S]=\int_0^{2\pi}\frac{\di\theta}{2\pi}R_z(\theta)\rho_S R_z(\theta)^\dag$ is dephasing noise on $\rho_S$.

\begin{figure}[h]
\centering
\includegraphics[height=5cm]{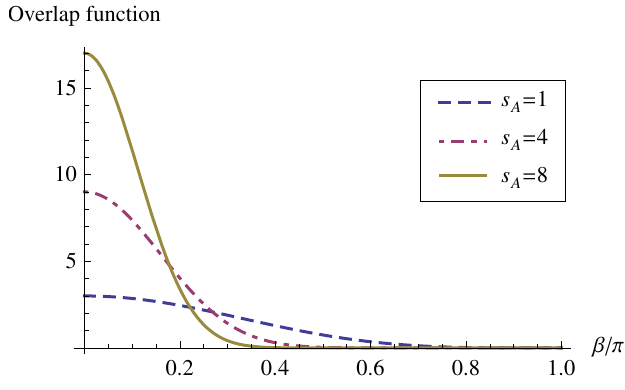}
\caption[Overlap function in the $SU(2)$ coherent state decoherence map (colour)]{(Color online) Overlap functions in $\beta$ for the $SU(2)$ coherent state decoherence map \eqref{eq-SU2CSF}, where $\beta=0$ to $\pi$ is the standard polar angle. The overlap function depends on $\beta$ and is unconstrained in $\alpha$ and $\gamma$: they are rotationally symmetric on the two-sphere about the north pole (the identity), indicating that there is mixing around circles of constant latitude. The distribution becomes more tightly peaked near $\beta=0$ as $s_A$ increases.\label{fig-mixedSU2fCS}}
\end{figure}

This overlap function is plotted in \figref{fig-mixedSU2fCS} for choices of $s_A$. Note that the function is rotationally symmetric about the direction $\beta=0$ on the two-sphere.
Indicating that there is only a single relative parameter: the angle between the two spins \cite{BRS04s}. Note that for this example the recovery map and therefore the change of reference frame procedure \eqref{eq-mainmeas} is not the optimal transformation, as defined in \secref{sec-REOptimal}. The optimal recovery map is a projection of $\m G_{SA}(\rho_S\otimes\rho_A)$ to total angular momentum $J$ \cite{BRS04s}.

\subsubsection{Classical limits of reference frame states}

We focus on the differences in the classical limits of this coset space example to the previous group examples. From \cite{RowedeGuiseSanders,BRSandersT}, for large $j$, the overlap function \eqref{eq-su2csoverlap} can be approximated as
\begin{equation}
\abs{_{\rm CS}\!\bk{j;e}{j;(\alpha,\beta,\gamma)}_{\rm CS}}^2=\cos^{4j}(\beta/2)\to \Ee^{-j\beta^2/2}.\label{eq-su2csoverlapdeltalim}
\end{equation}
This distribution has a variance of $\sigma^2=1/2j$. Up to normalisation we thus have that in the $j\to\infty$ limit $\exp[-j\beta^2]$ approaches the delta function $\delta(\beta)$. Therefore, for the $s_A\to\infty$ limit, the overlap in the decoherence map \eqref{eq-SU2CSF} becomes essentially a delta function in $\beta$. Then, although other examples of this scenario in this limit indicate no mixing, for coherent state reference frames we instead have that the decoherence function $\m F_S^{(A)}=\m D_S[\rho_S]$ is dephasing noise on $\rho_S$.  The final state then has a uniform average over $z$-rotations of the system state $\rho_S$, i.e.
\begin{equation}
\sigma_{S B}^h=\m E_{{\ket h_{\rm CS}}_B}\left[\m D_S(\m U_S(a^{-1})[\rho_S])\right].
\end{equation}
As this limit takes the form of an encoding from classical frame $A$ to quantum frame $B$, it demonstrates that the direction indicator reference frame fundamentally cannot encode phases. Interestingly, the $a$ and $h$ rotations do not commute with the dephasing operator, so we cannot write this in the usual form as $\m E_{{\ket{ah}_{\rm CS}}_B}(\m D_S[\rho_S])$.

When $s_B\to\infty$ the decoherence map is unaffected and we have an infinite size $B$ reference frame, which indicates the decoherence that would occur due to recovering the state $\rho_S$  from the reference frame $A$. Even in the simultaneous limit $s_A,s_B\to\infty$ there is still dephasing noise. The $U(1)$ dephasing is merely an artifact of describing the $SU(2)/U(1)$ coset in a representation of $SU(2)$.

\section{Conclusions\label{sec-conclusion}}

In this paper we investigated how the description of a state changes under a change of quantum reference frame in a static scenario. We did this by constructing a quantum operation which changes the quantum reference frame used to define a basis for another quantum system. We found that decoherence is in general induced on the quantum system due to the procedure. This decoherence is interpreted as a form of intrinsic decoherence due to a change of reference frame if one treats the frames as background parameters which possess fundamental quantum uncertainties. Our results may provide insight into what form a relativity principle would take in such a scenario. A relativity principle would dictate how the descriptions of a physical system and its dynamics change upon a change to a new quantum reference frame. This is distinct to the `equivalence principle' as studied in \cite{AharonovKaufherr}, where the choice of reference frame had an effect on relational measurements that used the reference frame; i.e.\;no active change of quantum reference frame was made. Examples of the change of quantum reference frame procedure for $U(1)$ and $SU(2)$ reference frames were presented.

\section*{Acknowledgements}
MP thanks Maki Takahashi, Terry Rudolph, Peter Turner, David Jennings, Hans Westman, Iman Marvian, Leon Loveridge and David Poulin for input and helpful discussions. We thank an anonymous reviewer for helpful suggestions. We acknowledge support from the ARC via the Centre of Excellence in Engineered Quantum Systems (EQuS), project number CE110001013.



%


\appendix
\section{Balanced Homodyne Detection of quantum phase references\label{app-BHD}}

In this section we will make some connections of relational quantum measurements with experiment. Balanced Homodyne detection is a measurement technique in quantum optics in which two beams are incident on either side of a beamsplitter. The angle of incidence with the plane of reflection is $45^\circ$ so that reflected and transmitted beams are on two paths, but these cannot mix with the incident beams. The beams on the two transmission paths are then measured with photon counters, returning numbers of photons $n_A$ and $n_B$. Therefore the projected state is a simultaneous number eigenstate for each path. It has total photon number $2j=n_A+n_B$, and difference in photons $2m=n_A-n_B$ where $m=-j$ to $j$ in integer steps \cite{TycSanders}. The basic idea is that the outcome $m/j$ is related to the relative phase of the two beams. If $j$ is large, there are more outcome possibilities, admitting a greater resolution of relative phase.

We want to see whether balanced homodyne detection is a way to perform the POVM \eqref{eq-M} that measures relative orientation of the reference frames $A$ and $B$. If it is, it provides an immediately experimentally accessible way to study the change of quantum reference frame procedure for phase references.  Indeed, there exists a coherent state amplification scheme using balanced homodyne measurement \cite{Josse06}, which may be considered as a specific change of quantum phase reference operation, from one coherent state to an amplified coherent state.

The standard treatment of balanced homodyne detection is that one input is the quantum state with a phase to be measured, and the second input is a classical `local oscillator', providing the phase reference for the measurement \cite{TycSanders}. In the scenario suggested in this section, whereby two quantum phase references are directly measured in a single measurement, we would need to consider the general situation where each input state is of finite size. Also we want to analyse the possibility of measurement of two phase eigenstates, as well as two coherent states. Since this view of a balanced homodyne detection treats both input beams equally as quantum states, the interpretation of the measurement is then that it measures relative phase of the two optical states, and requires no phase reference to do so. Adapting results in \refcite{TycSanders} we can analyse the large $s_A$ and $s_B$ limits of balanced homodyne measurements of coherent states and phase eigenstates. See also \refcite{VogelGrabow93}.

\subsubsection*{Two coherent states}
From \refcite{TycSanders} we have that the probability of $2j$ total photons and $2m$ difference in photons for two coherent states $\ket{s_A;a}_{\rm CS}$ and $\ket{s_B;b}_{\rm CS}$ is
\begin{multline}
P_m^j=\Ee^{-s_A^2}\Ee^{-s_B^2}\frac{1}{(j+m)!(j-m)!}2^{-2j}\times\\
\abs{s_A\Ee^{\ii a}-s_B\Ee^{\ii b}}^{2(j+m)} \abs{s_A\Ee^{\ii a}+s_A\Ee^{\ii b}}^{2(j-m)}.
\end{multline}
For two coherent states with equal amplitude $s$ and with phases $a$ and $b$, the measurement probabilities for $m$ are a function of $\cos^2[(b-a)/2]$:
\begin{multline}
P^j_m=\Ee^{-2s}\frac{(2s^2)^{2j}}{(2j)!}\binom{2j}{j+m}\times\\\left[\cos^2\left(\frac{b-a}2\right)\right]^{j+m} \left[1-\cos^2\left(\frac{b-a}2\right)\right]^{j-m}.
\end{multline}
The magnitude of the relative phase is monotonically mapped to $m\in[-j,j]$. Larger $j$ gives better relative phase accuracy. As $s\to\infty$, the resolution becomes perfect.

From \cite{TycSanders} we have for large coherent state $\ket{s_A;a}_{\rm CS}$ the outcome probability
\begin{equation}
P_m^j=\frac{\Ee^{-(2j-s_A^2)^2/2s_A^2}}{\sqrt\pi s_A^2} \abs{\Bigbk{x=\frac m{\sqrt j}}{\psi(b-a-\pi)}}^2\label{eq-appxP}
\end{equation}
where $x$ are the eigenvalues of $\hat x=(\hat a+\hat a^\dag)/\sqrt 2$, so for two coherent states with one amplitude $s_A$ large we replace $\ket{\psi(b)}\to\ket{s_B;b}_{\rm CS}$ and use the position representation of a coherent state \cite[Ch.V]{cohen1977quantum} to obtain
\begin{multline}
P^j_m=\frac{\Ee^{-(2j-s_A^2)^2/2s_A^2}}{\sqrt\pi s_A^2}(\pi\hbar)^{-\half}\times\\\exp[-\left(m/\sqrt j-s_B\cos(a-b+\pi)\right)^2].
\end{multline}
Again, this maps $\cos(b-a)$ to $m$. Probability $P^j_m$ is sharply peaked about $j=s_A^2/2$ for $s_A$ large, so we obtain accurate phase measurement.

\subsubsection*{Coherent state and phase eigenstate}
Balanced homodyne detection worked well as a phase measurement for a coherent state $\ket{s_B,b}_{\rm CS}$ with a coherent state $\ket{s_A,a}_{\rm CS}$ (treated in this case as the `reference' oscillator) because the state is localised in the $x$-$p$ phase space (with a Gaussian probability distribution). If we instead were measuring a phase eigenstate $\ket{\psi(b)}=\ket{s_B;b}$ and a large coherent state $\ket{s_A;a}_{\rm CS}$, we can again use \eqref{eq-appxP}. Calculating $\ex x$ and $\Delta x$ for the phase eigenstate using $\hat x=(\hat a+\hat a^\dag)/2$ on Fock states \cite[Ch.V]{cohen1977quantum}, we have in the large $s_B$ limit
\begin{align}
\ex x=&D_{s_B}^\half\frac23\cos b\qquad\text{and}\\
(\Delta x)^2\approx&\half\left(\frac32+\frac29D_{s_B}\cos^2b\right).
\label{eq-BHDPEvar}
\end{align}
As $B$ increases, the position variance is predominately determined by the phase $(b-a)$ of the state, and by the size of the Hilbert space on which the state has support, $D_{s_B}=s_B+1$. By \eqref{eq-appxP} we have a mapping of $\cos(b-a)$ to $m$.

\subsubsection*{Two phase eigenstates}

Consider balanced homodyne detection of two phase eigenstates $\ket{s_A;a}$ and $\ket{s_B;b}$ with size $s_A$ and $s_B$. The beamsplitter does not change total photon number probability. Therefore the total probability of detection of $2j$ photons is
\begin{multline}
P^j=(s_A+1)^{-1}(s_B+1)^{-1}\times\\(\min\set{j,s_A-j}+\min\set{j,s_B-j}+1).
\end{multline}
The probability grows from $D_{s_A}^{-1}D_{s_B}^{-1}$ at $2j=0$ linearly with $j$ to a plateau of $(\max\{s_A,s_B\}+1)^{-1}$ at $2j=\min\{s_A,s_B\}$ to $\max\{s_A,s_B\}$, then falling linearly with $j$ until probability is zero at $2j=s_A+s_B+1$. This plateau at moderate $j$ yields a lower average measurement accuracy than with two large coherent states. Modifying the derivation in \refcite{TycSanders} that led to \eqref{eq-appxP}, the probability for $\ket{j,m}$ of BHD of two phase eigenstates is approximately in the form of an overlap of a position eigenstate with a phase eigenstate:
\begin{multline}
P^j_m\approx D_{s_A}^{-1}D_{s_B}^{-1}\times\\\abs{j^{-\frac14}\bra{x=\frac m{\sqrt j}}\left(\sum_{k=\max\set{2j-s_A,0}}^{\min\set{s_B,2j}} \Ee^{\ii k(a-b-\pi)}\ket{k}\right)}^2.
\end{multline}
For $2j>s_A,s_B$, this superposition in the overlap looks like the $(s_B-s_A+2j+1)$ highest photon number components of a phase eigenstate $\ket{s_B;a-b-\pi}$. From \eqref{eq-BHDPEvar} we saw that the $x$ variance and expectation value of phase eigenstates depends on their cutoff and phase orientation. From our picture of phase eigenstates in phase space this overlap is the outer part of the pseudo-distribution of a phase eigenstate, without the inner part of the state. Since this becomes an isolated packet away from the origin, it allows some accuracy in correlation of a position measurement with the cosine of the phase. Notice however, as $j$ falls to $s_A/2$, the state becomes a complete phase eigenstate, so the overlap produces great inaccuracy mapping position to phase.

\subsubsection*{Balanced homodyne detection as a relational measurement}

In each case of study in this section, balanced homodyne detection provides a relational measurement of a parameter related to relative phase or quadrature. Since balanced homodyne detection projects to total photon number $j$ and a higher photon number offers more possible outcomes $m$, large coherent states are better suited to this type of measurement than phase eigenstates. Although the mapping between $m/j$ and relative phase is not linear, it becomes infinitely well resolved as the size of phase eigenstates or coherent states goes to infinity. However, only the relative phase modulo $\pi$ is measured. A final remark on the viability of balanced homodyne detection for use in the change of quantum reference frame procedure is that the homodyne detection destroys the state by absorbing the photons, so we would not be able to continue with the remainder of the change of reference frame procedure. Perhaps an extended optical setup could be utilised to produce an output state, similar to the balanced homodyne coherent state amplification scheme by \citet{Josse06}.

\end{document}